          [1999/12/02 1.01 (PWD)]
\documentclass[a4paper,twocolumn]{esapub} % European paper
\usepackage{natbib}
\usepackage{graphicx}
\usepackage{float}

\title{INTEGRAL  capabilities for faint Gamma-Ray Bursts}
\author{J.  Gorosabel}
\author{N. Lund}
\author{S. Brandt}
\author{N. J. Westergaard}
\affil{Danish Space Research Institute, Juliane Maries Vej 30, DK 2100 Copenhagen \O.}

\begin{document}


\maketitle

\begin{abstract}
  
  We discuss the INTEGRAL capabilities to detect a high redshift population
  of gamma-ray bursts (GRBs).  First a simple comparison between other past
  or planned experiments (BATSE, SAX, HETE-2, R$\o$mer, Swift) and INTEGRAL
  instrumentation (IBIS, JEM-X) is shown.  After this first view we will be
  focused on comparing the capabilities  of the two most sensitive missions
  (INTEGRAL/IBIS and Swift) of detecting a further  population of GRBs.  We
  conclude that,  if  the GRB  rate  is proportional to  the star formation
  rate, the capabilities of studying GRBs of  INTEGRAL are complementary to
  the ones of missions  like Swift and HETE-2,  specially devoted to prompt
  localizations of  GRBs.  Whereas Swift and  HETE-2 would detect  a higher
  number of GRBs than INTEGRAL ($\sim 8$ and $\sim 22$ more detections than
  IBIS and  JEM-X respectively), INTEGRAL  and specially IBIS  would detect
  very high redshift ($z  > 15$) GRBs,   unreachable for Swift and HETE-2.  
  This fact is very relevant for studying a population of GRBs further than
  $z=15$, which would be associated with the population {\rm III} of stars.
  Therefore, the INTEGRAL mission (precisely IBIS)  will be a very valuable
  tool to trace the primitive star formation rate at the early universe.

\end{abstract}

\section{Introduction}
The study of gamma-ray bursts (GRBs) is just one of  the many objectives of
INTEGRAL.  But developments  in the GRB-field  over the past few years have
made it  increasingly   clear that INTEGRAL   may  make a very  significant
contribution to this  fast developing field.  We  know today that most GRBs
originate  in the very distant  universe.  In fact,  we  believe that their
intrinsic      brightness  allows us to    detect    these events at epochs
corresponding to  the formation of  the earliest stellar populations.  Thus
they may  be  used as probes into  the  first stages of  star formation and
their spectra   may  reveal  the early    heavy-element  enrichment of  the
interstellar  medium.   Lamb \&  Reichart  (2000) claim that  GRBs could be
originated by primitive stellar populations up to redshift $z \sim 50$.

\section{The needs of the GRB community}
The discovery of the GRB afterglow, in X-rays, radio and the optical (Costa
et al. 1997, van  Paradijs et al. 1997) have  finally provided us the  long
sought tool  for associating a burst   with a concrete  object  in the sky. 
Although the duration of the afterglow is measured in hours or days instead
of the seconds which are characteristic of the bursts themselves, there are
still  two mandatory    requirements  for successful   afterglow  searches:
accurate  initial positions  and rapid dissemination   of the alerts.   The
successes   achieved by the SAX  team   have been based   on the  positions
accurate to maybe 10 square arcminutes and delays of a few hours (Boella et
al.   1997).  Lately, several afterglows have  also  been detected based on
arcminute positions provided   by the IPN  with delays  of the  order of 24
hours (Hurley et al.  2000a, 2000b). No  doubt, IPN positions will continue
to be useful and warrant follow up for a number of years to come, but if we
want to fully exploit the potential of high  resolution spectroscopy of the
intense phases  of the afterglows -  and if want  to check in higher detail
our  models  for  the afterglow  process   itself  - then  we  must provide
arcminute positions with delays of only a few minutes or even less.

\section{INTEGRAL capabilities}

INTEGRAL  will  be the  first gamma-ray  spacecraft which  combines imaging
instruments of high precision  and a continuous  real time telemetry link.  
In  Table 1 we  compare in a simple minded  way a number of different space
missions with capabilities for  GRB research.  The  missions are divided in
three groups depending on their energy range.  In the  first group the {\rm
  INTEGRAL/IBIS} sensitivity  is normalized respect  to Swift  sensitivity,
and in the second group the {\rm INTEGRAL/JEM-X}  and {\rm SAX} sensitivity
is given respect to the {\rm HETE-2} one.  For  consistency we have included
{\rm R$\o$mer/WATCH} and {\rm CGRO/BATSE} in a third group.

\begin{table}[t]
\caption{Capabilities~of~several~missions.}
\begin{tabular}{lcccccc}

\hline
Mission/instrument      &  Area    &  Coverage          &Relative &Energy  &   Orbit     & GRBs detected\\
                        &  cm$^2$  &  \% of $4\pi$ str. &sensitivity &  (keV)   &   efficiency& per year \\

\hline
\hline

Swift & 5200         &16       &   1.0  &  15 to 150& 0.6 &300 \\

INTEGRAL/IBIS(ISGRI)&  3000 &   1 &   3.0  &  15 to 150& 0.8 &35  \\   

\hline
\hline

HETE-2&   360 &    12&  1.0 & 2 to 25 &  0.5  &    25       \\

INTEGRAL/JEM-X&  1000& 15.4 & 4.0 & 2 to 25& 0.8 &  2 \\   

SAX/WFC   & 530($\times$2)     &2 ($\times$2) &  3.0 & 2 to 30 &  0.5 & 12 \\

\hline
\hline
                                                   
R$\o$mer/WATCH &95($\times$ 4)&25 ($\times$4)& 1.0&6 to 100&0.6 &70  \\

CGRO/BATSE & 2000 ($\times$8) & 65 & 4.3  &  50 to 300 &  0.6 & 300  \\

\hline                                                   
\end{tabular}

\end{table}

For calculating a value of the instrumental sensitivity (hereafter named as
$P_{ins}$) for each  experiment, we assume  that the sensitivity to a burst
is proportional  to the square-root   of the detector area,  and  inversely
proportional  to the square-root of   the background countrate.  Therefore,
this simple formula  becomes; $S \sim \sqrt{A/\Omega}$.   where  $S$ is the
sensitivity, $A$ is the detector area and $\Omega$ is the sky coverage. The
most reliable comparison can be done from Table 1 is between Swift and IBIS
(the ISGRI part)  because these are  very  comparable detector technologies
(CdZnTl in Swift, CdTl in IBIS).  Then assuming a sensitivity value of 0.04
ph  s$^{-1}$cm$^{-2}$  for Swift  (Gehrels 1999),  we   obtain  a 0.013  ph
s$^{-1}$cm$^{-2}$ sensitivity for IBIS.  HETE-2   and JEM-X share the  same
energy range  so one could  try to  obtain the  sensitivity of  JEM-X.  So,
considering a sensitivity  of 0.2  ph  s$^{-1}$cm$^{-2}$ for  \hbox{HETE-2}
(Ricker 1998)  and  the relative  sensitivity of   JEM-X respect to  HETE-2
(displayed    in  Table  1),   we    obtain  a   sensitivity of   0.012  ph
s$^{-1}$cm${-2}$ for  JEM-X.  However, their  different detector technology
make of this  number a  preliminary calculation of  the JEM-X  sensitivity. 
Besides the reduced number  of bursts that JEM-X  will detect ($\sim 2$ per
year) does not make  worthwhile to perform a  specific calculation in order
to study the JEM-X capabilities for faint bursts.  It is evident from Table
1 that IBIS on INTEGRAL will be the most sensitive gamma-ray burst detector
ever flown and not likely to be matched (excluding JEM-X which is detecting
a  few number of  them), sensitivity wise, by any  other mission within the
coming decade.  With this mission we have a unique opportunity!

\section{ Detectability of a faint population of GRBs}

We have selected  the most sensitive  future missions (INTEGRAL/IBIS, Swift
and HETE-2)  to calculate their capabilities  of  detecting a high redshift
population  of bursts. For the  estimate of the   number of GRBs that these
missions will detect, we assume:

\begin{itemize}

\item GRB spectra are power-laws; f$_{\nu} \sim \nu^{-\alpha}$

\vspace{12.0cm}

\hbox{}

\vspace{4.0cm}

\item The GRB rate is proportional to the star  formation rate (SFR) in the
  universe. The SFR  considered are the  one given by Rowan-Robinson (1999)
  for z  $<$ 5 and the  one calculated by Gnedin \&  Ostriker (1997)  for z
  $\ge$ 5. (See Fig.1).

\begin{figure}[H]
 \centering 
%   \resizebox{\hsize}{!} {\includegraphics{fig1.ps}}
   \resizebox{\hsize}{!} {\includegraphics{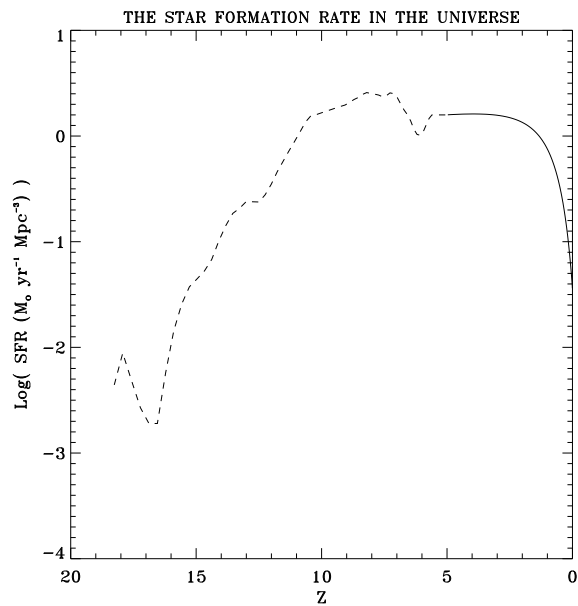}}
\caption{\label{Fig1} The plot shows  the Star formation rate (SFR) in
  the  universe as a function of  the redshift.  The dashed line represents
  the SFR derived by numerical simulations by Gnedin \& Ostriker (1997) for
  $z \ge 5$.  The solid line shows  the SFR at the $z  < 5$ region based on
  Observational estimates (Rowan-Robinson 1999). The transition between the
  two regions have been smoothed.}
\end{figure}

\item The GRB Luminosity function is given by;

\[ 
 S(L)=
  \left\{ \begin{array}{ll}
           L^{\beta}
      & \mbox{  $L_{min} < L < L_{max}$} \\
    0 & \mbox{  Otherwise } 
  \end{array} \right.
\]

being $L$ the peak     photon luminosity and $\beta$  the   luminosity
function index.  $L_{min}$, $L_{max}$  determine   the width of    the
luminosity function.

\item Although the effect of  several universe models  have been tried, the
  cosmological parameters  presented  in  this paper   are $\Omega_m$=0.3,
  $\Omega_{\Lambda}$=0.7, H$_{o}$ = 65 km s$^{-1}$ Mpc$^{-1}$.

\end{itemize}

According  to the  former assumptions  the differential  GRB detection
rate  at   a  given  photon  peak   flux  $P$  at   the  detector  (ph
cm$^{-2}$$s^{-1}$) is given by the following convolution integral;

\begin{equation}
 N_{GRB}(P) = C ~\Omega~e  \int_{0}^{\inf} R_{GRB} S(L) dL 
\end{equation}

$e$ is the efficiency of the orbit, $\Omega$ is  the instrumental coverage of
the  sky and  $R_{GRB}$ is the  GRB detection  rate  if  they were standard
candles,      i,e:   $R_{GRB}=     \frac{SFR(z)}{(1+z)}    \frac{dV(z)}{dz}
\frac{dz}{dP}$, being  $V$  the  comoving  volume  and $SFR(z)$  the   star
formation rate.  The value of the proportionality constant  $C$ is unknown. 
Fig 2.  shows $N_{GRB}(P)$ as  well as the  detection thresholds of several
instruments.

\begin{figure}[H]
\centering 
   \resizebox{\hsize}{!} {\includegraphics{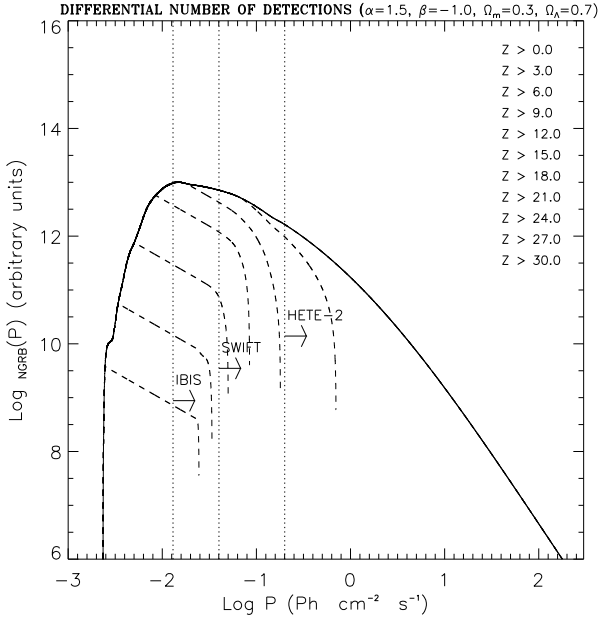}}
\caption{\label{Fig2} Differential peak photon flux distribution of GRBs.
  The  solid curve shows the  differential peak photon flux distribution if
  all redshifts are considered, i.e.  $N_{GRB}(P |  0)$.  The dashed curves
  represent the  differential peak  photon  flux distribution of  GRBs when
  only GRBs with $z > z_{edge}$ are taken  into account, i.e.  $N_{GRB}(P |
  z_{edge})$.  The  vertical  lines represent the detection  thresholds for
  the different instruments,  showing the arrows  the detectability region. 
  }
\end{figure}

 The  relationship  between $L$,  $z$ and  $P$   is given by  the next
expression;
$$  P= \frac{L}{ 4 \pi D(z)^2 (1+z)^{\alpha} }$$

Where $D(z)$  is the comoving distance. In  our calculations different
values of  $\alpha$, L$_{min}$, L$_{max}$ and  $\beta$ are considered.
The values  of $\alpha$, $\beta$,  $\Omega_{\Lambda}$ and $\Omega_{m}$
do not change  the final result qualitatively.  Instead  the values of
$L_{min}$,  $L_{max}$ are very  relevant to  the determination  of the
number  of  high  redshift  GRB  detections.   We  consider  the  most
pessimistic   case    where   $L_{min}=10^{57}$   ph    s$^{-1}$   and
$L_{max}=10^{58}$ ph s$^{-1}$ (according to the GRB redshifts measured
so  far  the GRB  luminosity  function seems  to  be  wider).  We  can
calculate  the contribution  to  the  integral (1)  by  the GRBs  with
redshift larger than $z_{edge}$;

{\small $$N_{GRB}(P | z_{edge}) = C~\Omega~e \int_{0}^{\inf} H(z(L),z_{edge}) R_{GRB} S(L)~dL$$}

Where $H(z(L),z_{edge})$ is a step function that vanishes unless $z(L)
> z_{edge}$.    Obviously,  $N_{GRB}(P)   =  N_{GRB}(P   |   0)$,  and
$\frac{N_{GRB}(P  | z_{edge}) }{ N_{GRB}(P)} \le  1$. Finally  we can
calculate  the number  of  GRBs detected  above  a given  instrumental
photon  flux  threshold  $P_{ins}$  that have  redshifts  larger  than
$z_{edge}$;

$$N_{GRB}(z_{edge}|P_{ins})=\int_{P_{ins}}^{\inf}  N_{GRB}(P | z_{edge}) dP $$

The ignorance of the proportionality constant  $C$ prevents us to derive an
absolute value for  $N_{GRB}(z_{edge}|P_{ins})$.  However, we can determine
the            relative   quantity         $\frac{N_{GRB}(z_{edge}|P_{ins})
  }{N_{GRB}(0,P_{ins})}$, which  provide us  the  proportion of  detections
that have a redshift larger than $z_{edge}$ (see Fig. 3).

\begin{figure}[H]
\centering
   \resizebox{\hsize}{!} {\includegraphics{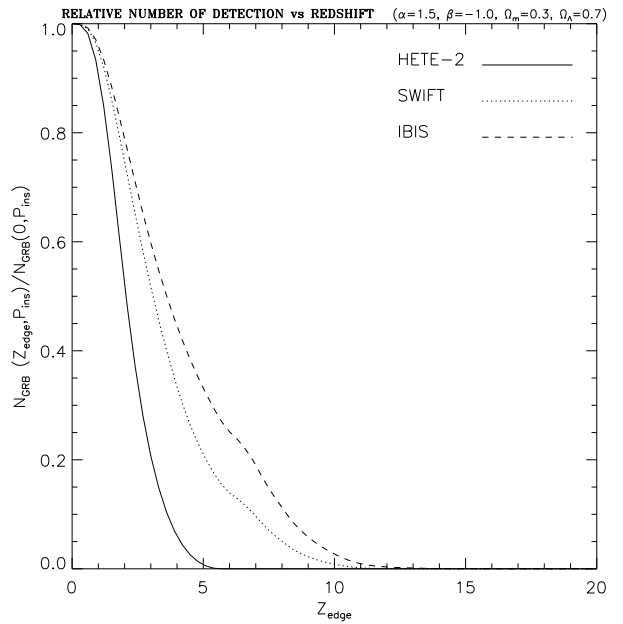}}
\caption{\label{Fig3} Relative  number of detections as  a function of
  the  redshift.  This   plot shows  for  several missions/instruments  the
  fraction  of the  detected  GRBs  that   have   a redshift  larger   than
  $z_{edge}$.}
\end{figure}

\section{IBIS {\scriptsize vs.} Swift; Comparison of the number of the GRB detections}

As it is shown in Fig. 3, $\sim$10\% of the GRBs detected by IBIS will have
a redshift larger than 8.4.  For Swift the $z >8.4$ population will be just
$\sim$  4\% of  the  total number    of  detections.  HETE-2   is  the less
sensitivity detector, being constrained to detect GRBs with redshifts $ z <
6$.  Therefore we will  not consider HETE-2 for the  further study aimed to
calculating    the relative  number  of detections    as a  function of the
redshift. We will be centered  in comparing IBIS   and Swift capabilities.  
Besides, as we noted, the similar energy range and detector technologies of
IBIS and Swift guarantee a reliable calculation of this fraction.

For determining the relative number of  detections between two experiments,
$A$ and $B$, the next expression has to be calculated:

$f_{A/B}(z_{edge})={\small \frac{\Omega_A
    e_A\int_{P_{A}}^{\inf}\int_{0}^{\inf}
    H(z(L),z_{edge})R_{GRB}S(L)dLdP}{\Omega_B e_B \int_{P_{B}}^{\inf}\int_{0}^{\inf}H(z(L),z_{edge})R_{GRB}S(L)dLdP}}$

This  function will give  the relative number  of  GRB detections with $z >
z_{edge}$.  We  have applied the former  expression to  derive the fraction
$f_{IBIS/Swift}$ as a function of the GRB redshift.

If we consider  $z_{edge}=0$ the  fraction  $f_{IBIS/Swift}$ gives us   the
fraction of GRBs detected with $z>0$, i.e, all the detections independently
of    their  redshifts  are    considered.      We  obtain   a value     of
$f_{IBIS/Swift}(0)= 1/7.8$ (see Fig.~4), which is consistent with the value
of $f_{IBIS/Swift}$ derived from  last column of Table  1.  The large field
of  view (FOV) of  Swift in comparison  to IBIS makes  that for $z_{edge} <
11.6$ $f_{IBIS/Swift} < 1$.  Instead, for further redshifts than 11.6, IBIS
sensitivity becomes the governing factor and $f_{IBIS/Swift} > 1$.

\section{Conclusion}
Fig.    3 shows  $\frac{N(z_{edge}, P_{ins})}{N(0,P_{ins})}$,  for  HETE-2,
IBIS, and   Swift.  For  IBIS  the  tail  of $\frac{N(z_{edge},  P_{ins})}{
  N(0,P_{ins})}$  extends to redshifts  $z_{edge} >  11$.  Swift and HETE-2
would detect  a closer population    of burst, specially  HETE-2 would   be
constrained to redshifts $z_{edge} < 6$.

\begin{figure}[h]
  \centering \resizebox{\hsize}{!}  {\includegraphics{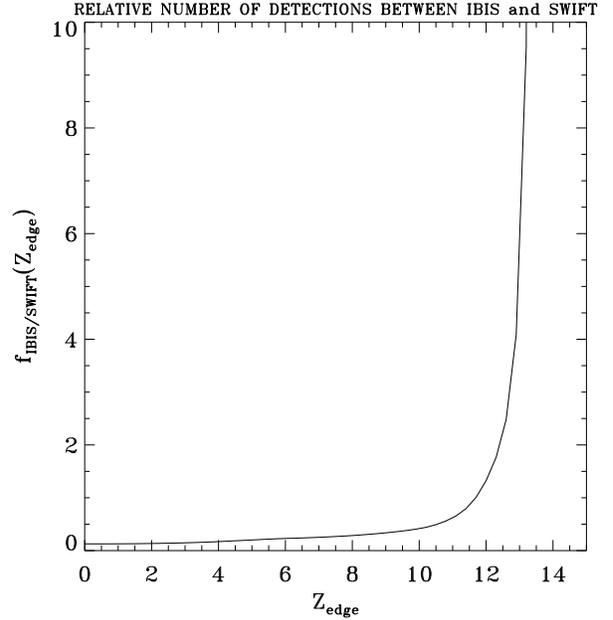}}  

\caption{The figure
  shows the GRB detection ratio between IBIS and Swift as a function of the
  the GRB population redshift.}
\end{figure}

If we   consider all  the GRBs   ($z_{edge} >  0$)  the detection  fraction
$f_{IBIS/Swift}$=1/7.8. Thus, at  low redshifts the  large FOV of  Swift in
comparison to INTEGRAL  instrumentation governs the  number of detections.  
However, at high  redshifts   the better sensitivity   of IBIS   makes  the
fraction of detections $f_{IBIS/Swift} >1$ (for redshifts $z_{edge} > 11.6$
$f_{IBIS/Swift} >1$, see Fig.  4).  Although JEM-X  FOV and sensitivity are
less suitable than the one of IBIS to detect GRBs, the spectral peak of the
high  redshift GRBs  (usually at  500--1200  keV) will be in  the detection
range of JEM-X.  Therefore JEM-X will be also a very valuable tool to study
the high redshift GRBs.

In conclusion, the capabilities of studying GRBs of JEM-X and IBIS on board
INTEGRAL are complementary  to the ones  of missions like  Swift and HETE-2
specially devoted  to  prompt  localizations of  GRBs.   Whereas Swift  and
HETE-2 would detect  more GRBs than   INTEGRAL, JEM-X and  IBIS instruments
would detect  very high redshift   GRBs unreachable to the above  mentioned
missions.  Therefore, INTEGRAL  and specially IBIS will  be a very valuable
tool to trace the SFR rate in the early universe.

\section*{Acknowledgments}

J. Gorosabel acknowledges the receipt of a Marie Curie Research Grant from
the European Commission.

% The following bibliography was produced with
%   \bibliographystyle{aa}
%   \bibliography{esapub}
% The results are inserted directly here to simplify
% the demonstration.

%\begin{table}[b]
%\begin{center}
%\caption{The table displays the GRB detection~capabilities for several missions.}
%\begin{tabular}{lcccccc}

%\hline
%\hline
%{\tiny Mission }   &{\tiny  Area} &{\tiny  Coverage} &{\tiny  Relative}  &  {\tiny Energy}  &{\tiny   Orbit} &{\tiny GRBs} \\
%             & {\tiny cm$^2$}&{\tiny  \% of $4\pi$}&{\tiny  sensitivity}&{\tiny  (keV)} &{\tiny   efficiency}&{\tiny per year} \\

%\hline

%{\tiny BATSE} & {\tiny 2000 ($\times$8)}&{\tiny 65} & {\tiny 5.6/1.0} &{\tiny  50-300} &{\tiny  0.6} &{\tiny 300}\\

%{\tiny SAX   }&{\tiny 530($\times$2)} &{\tiny 2 ($\times$2)} &{\tiny  16.2/2.9} &{\tiny 2 to 30} &{\tiny  0.5} &{\tiny 12} \\

%{\tiny HETE-2 }&{\tiny   360}&{\tiny   12}&{\tiny  5.5/1.0}&{\tiny 2 to 25}&{\tiny  0.5}&{\tiny    25}\\

%{\tiny INTEGRAL$^*$}&{\tiny 3000}&{\tiny   1}&{\tiny   54.8/9.8}&{\tiny  15 to 150}&{\tiny 0.8}&{\tiny  35}\\   
                                                   
%{\tiny R$\o$mer}&{\tiny 95($\times$ 4)}&{\tiny 25 ($\times$4)}&{\tiny 2.0/0.4}&{\tiny  6 to 100}&{\tiny 0.6}&{\tiny 300}  \\

%{\tiny Swift}&{\tiny 5200} &{\tiny  16} &{\tiny 18.0/3.2}&{\tiny 15 to 150} &{\tiny 0.6} &{\tiny 300}  \\
                                                   
%\hline

%*{\tiny (IBIS, ISGRI)} &&&&&&\\
%\hline
%\end{tabular}

%\end{center}
%\end{table}

\end{document}